\documentclass[aps,prl,twocolumn,superscriptaddress,showpacs]{revtex4}
\usepackage{graphicx}

\usepackage{dcolumn}

\usepackage{bm}
\usepackage{color}
\usepackage[colorlinks]{hyperref}
\input{epsf}

\begin{document}


\title{Zero-Resistance States Induced by Electromagnetic-Wave Excitation
in GaAs/AlGaAs Heterostructures}

\author{R. G. Mani}
\email{mani@deas.harvard.edu} \affiliation {Harvard University,
Gordon McKay Laboratory of Applied Science, 9 Oxford Street,
Cambridge, MA 02138, USA} \affiliation {Max-Planck-Institut
f\"{u}r Festk\"{o}rperforschung, Heisenbergstrasse 1,  Stuttgart
70569, Germany}

%
%
%
%
\date{June 15, 2003}
\begin{abstract}
We report the detection of novel zero-resistance states induced by
electromagnetic wave excitation in ultra high mobility GaAs/AlGaAs
heterostructure devices, at low magnetic fields, $B$, in the large
filling factor limit. Vanishing resistance is observed in the
vicinity of $B = [4/(4j+1)] B_{f}$, where $B_{f} =
2\pi\textit{f}m^{*}/e$, where m$^{*}$ is the effective mass, e is
the charge, and  \textit{f} is the microwave frequency. The
dependence of the effect is reported as a function of \textit{f},
the temperature, and the power.
\end{abstract}
%
\pacs{73.21.-b,73.40.-c,73.43.-f; Journal-Reference: Physica E
(Amsterdam) \textbf{22}, 1 (2004).}
%
\maketitle Vanishing electrical resistance in condensed matter has
introduced new physical phenomena such as superconductivity, which
developed from the detection of a zero-resistance state in a
metal.\cite{tinkham} More recently, the discovery of quantum Hall
effects (QHE) stemmed from studies of zero-resistance states at
low temperatures $(T)$ and high magnetic fields $(B)$ in the
2-Dimensional Electron System (2DES).\cite{prange,ando,tsui}
Quantum Hall effect and superconductivity have shown that a
complex electronic system can exhibit instantly-recognizable
physical phenomena. They have also demonstrated that observations
of vanishing resistance in unusual settings can be a harbinger of
new physics. Here, we report the observation of novel vanishing
resistance states in an unexpected setting - the ultra high
mobility 2DES irradiated by low energy photons, at low
temperatures, in the low magnetic field, large filling factor
limit. We find that GaAs/AlGaAs heterostructures including a 2DES
exhibit vanishing diagonal resistance about $B$ = $(4/5) B_{f}$
and $B$ = $(4/9) B_{f}$, where $B_{f}$ = $2\pi f m^{*}/e$, $m^{*}$
is an effective mass, $e$ is the electron charge, and $f$ is the
radiation frequency. And, the resistance-minima follow a series
$B$ = $[4/(4j+1)] B_{f}$ with $j$=$1,2,3$... Remarkably, in this
instance, vanishing resistance in the 2DES does not produce
plateaus in the Hall resistance, although the diagonal resistance
exhibits activated transport and zero-resistance states, similar
to QHE.\cite{maniBAPS,maninature}

\begin{figure}[b]
\begin{center}\leavevmode
\includegraphics[scale = 0.25,angle=0,keepaspectratio=true,width=3in]{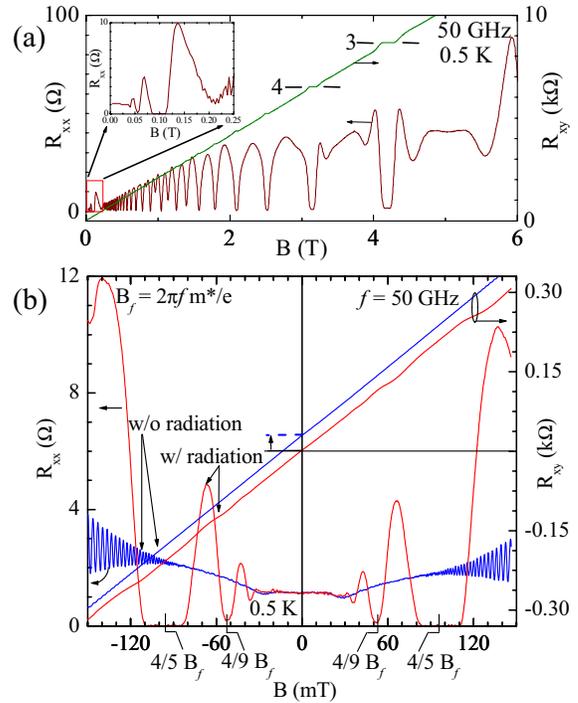}
\caption{(a): The Hall ($R_{xy}$) and diagonal ($R_{xx}$)
resistances in a GaAs/AlGaAs heterostructure with excitation at 50
GHz. Quantum Hall effects (QHE) occur at high B as $R_{xx}$
vanishes. Inset: An expanded view of the low-B data.  (b): Data
over low magnetic fields obtained both with (w/) and without (w/o)
excitation at 50 GHz. Here, radiation induced vanishing resistance
about $(4/5) B_{f}$ does not induce plateaus in the Hall
resistance, unlike in QHE. Yet, there are perceptible microwave
induced oscillations in the Hall effect.}
\label{figurename1}\end{center}\end{figure}

Measurements were performed on Hall bars, square shaped devices,
and Corbino rings, fabricated from GaAs/AlGaAs heterostructures.
After a brief illumination by a red LED, the best material was
typically characterized by an electron density, $n$(4.2 K)
$\approx$ $3$ $\times$ $10^{11}$ cm$^{-2}$, and an electron
mobility $\mu$(1.5 K) $\approx$ $1.5$ $\times$ $10^{7}$
cm$^{2}$/Vs. Lock-in based four-terminal electrical measurements
were carried out with the sample mounted inside a waveguide and
immersed in pumped liquid He-3 or He-4. Electromagnetic (EM) waves
in the microwave part of the spectrum, $27 \leq f \leq 170$ GHz,
were generated using various tunable sources. The power level in
the vicinity of the sample was typically $\leq$ 1 mW. In this
report, we shall mainly exhibit data which illustrate the
characteristics of the phenomena. Expanded discussions appear
elsewhere.\cite{maninature,manibeats,maniICPS,maniIV}

\begin{figure}[t]
\begin{center}\leavevmode
\includegraphics[scale=0.25,angle=0,keepaspectratio=true,width=3in]{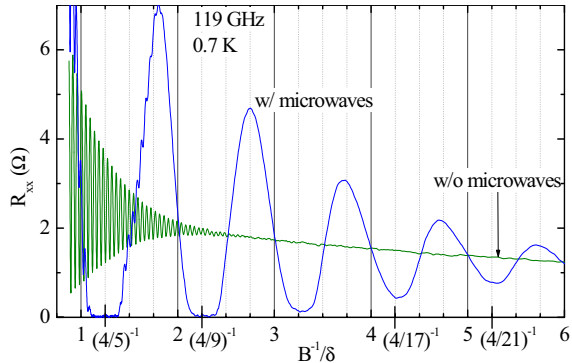}
\caption{ A plot of $R_{xx}$ versus the normalized inverse
magnetic field, $B^{-1}/\delta$, demonstrates periodicity of the
radiation induced oscillations . The curves with- and without-
radiation intersect near integral and half-integral values of the
abscissa.} \label{figurename2}
\end{center}
\end{figure}

Fig. 1 (a) shows measurements of the diagonal ($R_{xx}$) and Hall
$(R_{xy})$ resistances to $B$ = 6 Tesla where, under microwave
excitation at 50 GHz, $R_{xx}$ and $R_{xy}$ exhibit the usual
quantum Hall behavior for $B$ $\geq$ 0.3 Tesla.\cite{prange,ando}
In contrast, for $B$ $<$ 0.25 Tesla, see inset of Fig. 1(a), a
radiation induced signal occurs and the resistance vanishes over a
broad B-interval about $B$ = 0.1 Tesla. Further high-resolution
measurements are shown in Fig. 1 (b). Without EM-excitation,
$R_{xx}$ exhibits Shubnikov-deHaas oscillations for $B$ $>$ 100
milliTesla (Fig. 1 (b)). The application of microwaves induces
resistance oscillations,\cite{maniBAPS,zudovprb,ye} the resistance
under radiation falls below the resistance without radiation, over
broad $B$-intervals.\cite{maniBAPS,maniICPS} Indeed, $R_{xx}$
appears to vanish about $(4/5)
B_{f}$.\cite{maninature,manibeats,maniIV} Although these
zero-resistance-states exhibit a flat bottom as in the quantum
Hall regime,\cite{prange} $R_{xy}$ under radiation does not
exhibit plateaus over the same $B$-interval.

As an example, Fig. 2 shows a normalized $B^{-1}$ plot of the
response obtained under microwave excitation at 119 GHz. This plot
shows that (i) the magnetoresistance oscillations are periodic in
$B^{-1}$, (ii) the minima form about $B_{min}^{-1}/\delta$ =
$[4/(4j+1)]^{-1}$ with $j=1,2,3$... Here, $\delta$ is the
oscillatory period in $B^{-1}$, which is consistent with
$B_{f}^{-1}$ within experimental uncertainty, (iii) the higher
order maxima coincide with $B_{max}^{-1}/\delta$ =
$[4/(4j+3)]^{-1}$. Experiment indicates that the resistance maxima
generally obey this rule for integral $j$, excepting $j$ = 0,
where phase distortion associated with the last peak seems to
shift it from the $B_{max}^{-1}/\delta$ = 3/4 to approximately
0.85 ($\pm$ 0.02) (see also Fig. 5), and (iv) the data obtained
with radiation cross the data obtained without radiation, at
integral and half-integral values of the $B^{-1}/\delta$, when
$B^{-1}/\delta$ $\geq$ 2. At these $B^{-1}/\delta$, the photon
energy $hf$ spans, perhaps, an integral, $j$, or half integral, $j
+ 1/2$, cyclotron energies. The crossing feature near integral
$B^{-1}/\delta$ looks to be in agreement with
theory,\cite{durst,shi&xie1} which suggests that, when $2\pi f
/\omega_{C}$ = $j$, the conductivity in the presence of radiation,
$\sigma$, equals the conductivity in the absence of radiation,
$\sigma_{dark}$, i.e., $\sigma$ = $\sigma_{dark}$.\cite{shi&xie1}
Remarkably, the data suggest the same behavior at $j$ + 1/2.

\begin{figure}[t]
\begin{center}\leavevmode
\includegraphics[scale = 0.25,angle=0,keepaspectratio=true,width=3in]{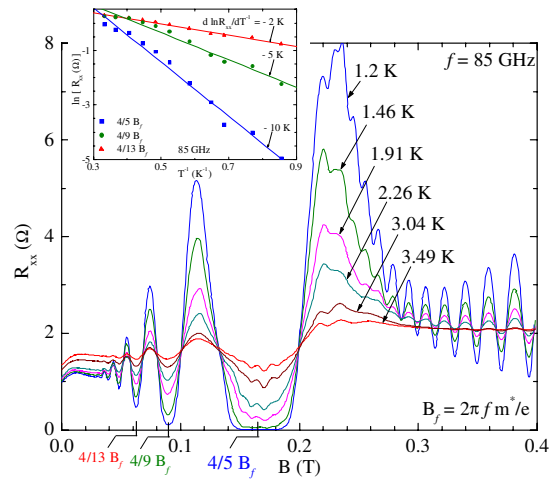}
\caption{The T-dependence of $R_{xx}$ at 85 GHz under
constant-power radiation. The radiation induced resistance minima
become deeper at lower temperatures. (inset) A $ln$ $R_{xx}$ vs.
$T^{-1}$ plot at $B = (4/5) B_{f}$, $(4/9) B_{f}$, and $(4/13)
B_{f}$ suggests activated transport. The activation energy $T_{0}$
$\approx$ 10 K for the $(4/5) B_{f}$ state at 85 GHz.}
\label{figurename3}\end{center}\end{figure}

The temperature variation of $R_{xx}$ at 85 GHz, shown in Fig. 3,
displays both the strong $T$-dependence of $R_{xx}$ and the low-T
requirement for the observation of the radiation induced
zero-resistance state. The T-variation of $R_{xx}$ at the deepest
minima suggests activated transport, i.e., $R_{xx}$ $\sim$
$\exp(-T_{0}/T)$,\cite{prange,maninature} see inset Fig. 3, and,
in temperature units, the activation energy at $B$ = (4/5)
$B_{f}$, $T_{0}$ $\approx$ 10 K at 85 GHz, exceeds the Landau
level spacing, 3.26 K, and the photon energy, $hf$ = 4.08 K.
Remarkably, $T_{0}$ is reduced as one moves to higher order minima
at lower $B$, see inset Fig. 3. Further studies indicated that
$T_{0}$ varied with the radiation intensity, as illustrated in
Fig. 4.

\begin{figure}[t]
\begin{center}\leavevmode
\includegraphics[scale = 0.25,angle=0,keepaspectratio=true,width=3.25in]{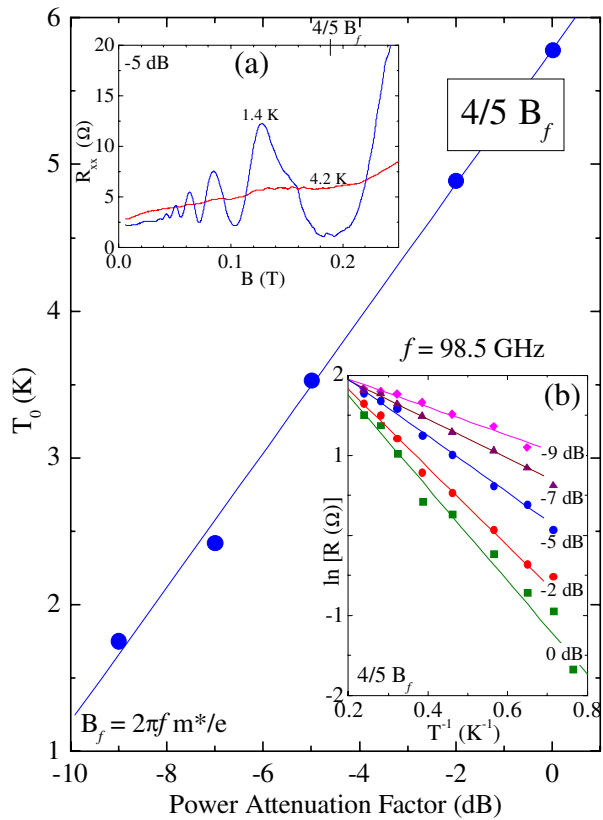}
\caption{The variation of the activation energy, $T_{0}$, as a
function of the microwave power at 98.5 GHz and $(4/5) B_{f}$. (a)
$R_{xx}$ vs. $B$ at an attenuation factor of -5 dB. (b) The
natural logarithm of $R_{xx}$ is shown versus the inverse
temperature.}\label{figurename4}\end{center}\end{figure}

These power dependent activation results go together with other
measurements, which showed that the amplitude of the radiation
induced resistance oscillations increased with the radiation
intensity (see Fig. 3a,ref\cite{maninature}), up to a
(sample-dependent) optimum value of the radiation intensity (see
Fig. 5). Concomitantly, the resistance in the vicinity of $B$ =
$[4/(4j+1)] B_{f}$ decreased and, about $(4/5) B_{f}$ and $(4/9)
B_{f}$, $R_{xx}$ $\rightarrow$ 0. Current dependence measurements
also demonstrated insensitivity to the magnitude of the current
and the Hall electric field.\cite{maninature,manibeats}

\begin{figure}[t]
\begin{center}\leavevmode
\includegraphics[scale = 0.25,angle=0,keepaspectratio=true,width=3in]{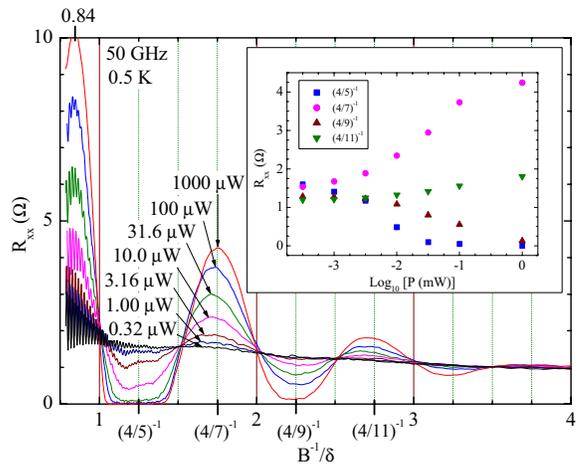}
\caption{Radiation induced resistance oscillations at $f$ = 50 GHz
are exhibited for various source intensities, in units of $\mu$W.
The inset shows the extrema resistance at (4/5)$^{-1}$,
(4/7)$^{-1}$, (4/9)$^{-1}$,  (4/11)$^{-1}$ vs. the logarithm of
the power. } \label{figurename5}\end{center}\end{figure}

The effect of tilting the specimen with respect the applied $B$,
as the microwave radiation is directed along the $B$-axis, is
illustrated in Fig. 6. Note that the field scale shown on the
abscissa is different, for each panel in the figure. These results
suggest that the radiation induced magnetoresistance, ignoring
possible role for spin, is mostly sensitive to the perpendicular
component of the total magnetic field, $B_{\perp} = B
\cos(\theta)$.

The orientation of the microwave polarization with respect to the
current axis could be an important factor in experiment
\cite{koulakov} and, therefore, results examining the role of this
parameter are shown in Fig. 7. Measurements carried out at 39 GHz
on a sample mounted inside a WR-28 waveguide (see Fig. 7), with a
current applied between the ends of the L-shaped device, suggest
the same period and phase in the $E$ $\parallel$ $I$ and the $E$
$\perp$ $I$ configurations.

Some results of an investigation examining the $I$-dependence of
the resistance in the Corbino geometry are exhibited in Fig. 8.
This figure shows that the Corbino resistance is independent of
the current, at relatively low currents, which supplements
previously reported results for the Hall geometry (see Fig. 3(b),
ref. \cite{maninature}).

\begin{figure}[t]
\begin{center}\leavevmode
\includegraphics[scale = 0.25,angle=0,keepaspectratio=true,width=3.25in]{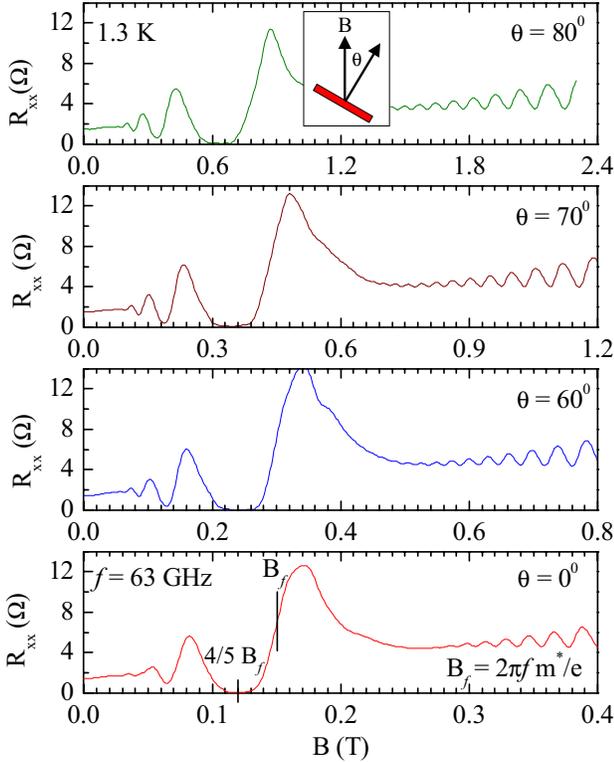}
\caption{Tilt-field measurements show that the radiation induced
magnetoresistance is sensitive to the perpendicular component of
total applied magnetic field.
}\label{figurename6}\end{center}\end{figure}

In this configuration, a maximum in the resistance (or a minimum
in the conductance) occurs about $B$ = (4/5) $B_{f}$ and $B$ =
(4/9) $B_{f}$, in place of the resistance minimum observed in the
Hall configuration (see Fig. 1). This geometry difference
originates from a well-known feature of transport, that the
resistance in the Corbino geometry $R_{C}$ $\approx$
$\sigma_{xx}^{-1}$, while in the Hall geometry, $R_{xx}$ $\approx$
$\sigma_{xx}/\sigma_{xy}^{2}$, under $\omega_{C}\tau$ $>$ 1
condition.

Preliminary results of  experiments examining possible
transmission and absorption characteristics under microwave
excitation are illustrated in Fig. 9 and Fig. 10.  For the
transmission experiments (Fig. 9), a carbon resistor was placed
immediately below the sample, and its resistance was measured
simultaneously along with the specimen characteristics, as
radiation was applied from above (see Fig. 9, inset). The
magnetoresistance of the carbon resistor is shown in Fig. 9(a),
while Fig. 9(b) illustrates $R_{xx}$ of the 2DES vs. $B$. Here,
-60 dB corresponds to vanishing excitation, while 0 dB corresponds
to the maximum microwave intensity. In this sample, the optimum
radiation induced $R_{xx}$ response, shown in Fig. 9(b), was
observed in the vicinity of -8 dB. That is, the amplitude of the
radiation induced oscillations increased monotonically with
increasing power attenuation factor up to -8 dB.
\begin{figure}[t]
\begin{center}\leavevmode
\includegraphics[scale = 0.25,angle=0,keepaspectratio=true,width=3.25in]{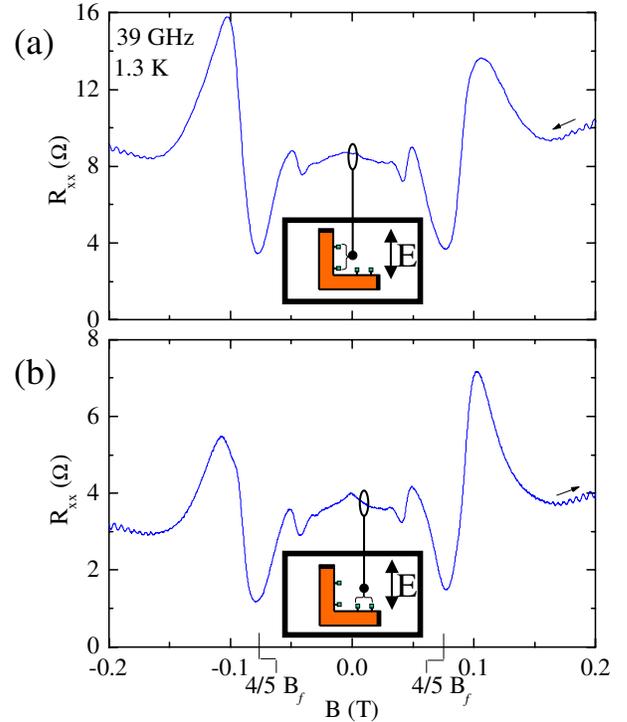}
\caption{Measurements carried out on an L-shaped sample mounted
inside a WR-28 waveguide suggest insensitivity to the E-field
polarization.(a) The current is oriented parallel to the E-field
axis. (b) The current is oriented perpendicular to the E-field
axis. }\label{figurename7}\end{center}\end{figure}

A further increase of the radiation intensity (dB $\rightarrow$ 0)
produced a reduction in the peak height along with an increase in
the resistance at the minima (see Fig. 9(b)). This behavior
suggests a "breakdown" of the radiation induced zero-resistance
states at high excitation levels, which is somewhat analogous to
the "breakdown" of the quantum Hall effect that is observed at
high currents.\cite{prange} At the same time, the response of the
carbon resistor placed below the sample (see Fig. 9(a)) suggests
structure at magnetic fields mostly above $B_{f}$, which becomes
more pronounced with increased excitation. The feature correlates
with a strong radiation-induced decrease in $R_{xx}$ just above
0.3 Tesla. Observed oscillations in $R_{xx}$ below $B_{f}$ (Fig.
9(b)) are imperceptible in the detector response (see Fig. 9(a)).

The results of a measurement reflecting possible absorption
characteristics of the specimen are illustrated in Fig. 10. In
this experiment, the sample was mounted inside a waveguide that
was susceptible to He-4 thermo-acoustic oscillations
(TAO).\cite{TAO}
\begin{figure}[b]
\begin{center}\leavevmode
\includegraphics[scale = 0.25,angle=0,keepaspectratio=true,width=2.5in]{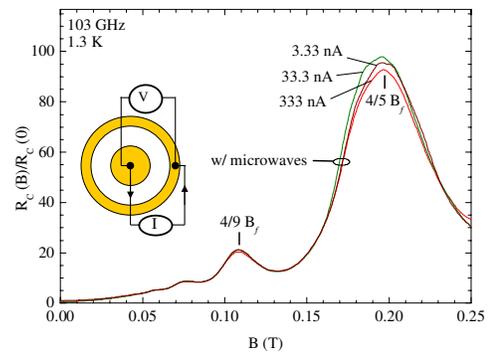}
\caption{In the Corbino geometry, resistance maxima occur in the
vicinity of $B = [4/(4j+1)] B_{f}$, where $j=1,2,3,...$ The
Corbino resistance, $R_{C}$, was found to be independent of the
current in the investigated
regime.}\label{figurename8}\end{center}\end{figure}

In this setup, the onset of TAO's under microwave excitation
generated mechanical vibrations of the sample and thereby produced
electrical noise in the measurement. However, this behavior was
only observed for $B$ $<$ $B_{f}$, which suggested the conjecture
that energy absorption by the sample at $B$ $<$ $B_{f}$ modified
the immediate sample environment and triggered the TAO's. That is,
in Fig. 10, the 'noisy' portion of the data could signify
absorption by the sample below $B_{f}$ (at 4.2 K).

Radiation induced zero-resistance states have been confirmed in
GaAs/AlGaAs quantum wells by Zudov et al.,\cite{zudovprl}, who
suggested the effect as evidence of a new dissipationless effect
in 2D electronic transport.

Phillips has suggested the effect to be a manifestation of sliding
charge density waves.\cite{phillips} Durst and co-workers have
identified radiation induced resistance oscillations with a driven
current, similar to Ryzhii.\cite{durst,ryzhii}

\begin{figure}[t]
\begin{center}\leavevmode
\includegraphics[scale = 0.25,angle=0,keepaspectratio=true,width=3.25in]{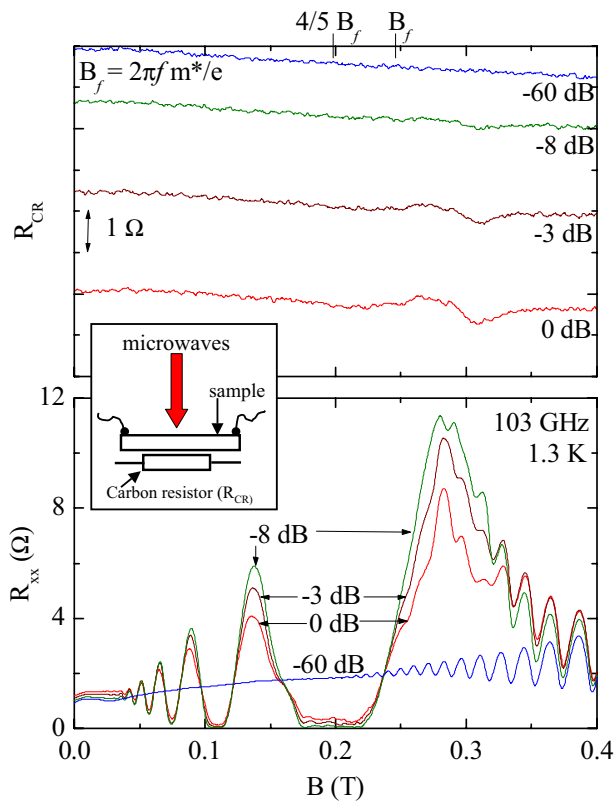}
\caption{These measurements examine the transmission
characteristics of the 2DES under irradiation. Here, a carbon
resistor placed below the sample served as the radiation detector.
The detector resistance, $R_{CR}$, vs. $B$ (top). $R_{xx}$ vs. $B$
of the 2DES (bottom). This shows that when the power attenuation
factor exceeds -8dB, the oscillation amplitude decreases
signifying "breakdown". The detector response suggests non
monotonic transmission above $B_{f}$, while strong features do not
occur below $B_{f}$ in top panel}
\label{figurename9}\end{center}\end{figure}

Andreev et al.,\cite{andreev} and Anderson and Brinkman,\cite{PWA}
suggested a physical instability for the negative resistivity
state, while Andreev et al. proposed a scenario for realizing
zero-resistance in measurement.\cite{andreev}  A complementary
theory by Shi and Xie,\cite{shi&xie1} also realized
magnetoresistance
oscillations.\cite{maninature,manibeats,maniICPS}.

\begin{figure}[t]
\begin{center}\leavevmode
\includegraphics[scale = 0.25,angle=0,keepaspectratio=true,width=2.75in]{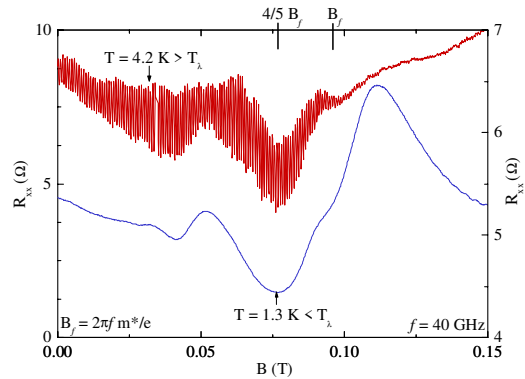}
\caption{Possible indication of energy absorption by the specimen
below $B_{f}$. In a helium-4 (He-4) cryostat, a sample was mounted
inside a waveguide, which was susceptible to He-4 Thermo-Acoustic
Oscillations(TAO). Above $\approx$ 2.2 K, and below $B_{f}$, TAO's
were observed, which produced sample vibrations, and therefore,
electrical noise in the measurement. Below the Lambda-point, this
noise producing mechanism vanished. The results suggest that
energy absorption by the sample triggered the TAO's in this
setup.} \label{figurename10}\end{center}\end{figure}

The observed zero-resistance states have also been related to a
quantum Gunn effect,\cite{shrivastava2} while Rivera and Schulz
have pointed out the possibility of gap formation due to the
replication of Landau levels in the presence of
radiation.\cite{riviera} Although there has been progress  in
understanding aspects, see ref. \cite{fitzgerald,klesse}, many
features including the activated temperature dependence, and the
zero-resistance states themselves, could be better understood.

We acknowledge stimulating discussions with J. H. Smet, K. von
Klitzing, V. Narayanamurti, and W. B. Johnson. The high quality
MBE material was kindly provided by V. Umansky. Special thanks to
K. von Klitzing for providing insight, support, and encouragement
over many years.

\end{document}